\documentstyle[12pt,epsfig]{article}

\begin{document}

\title{Analytical estimation of the Earth's magnetic field scale}

\author{Mauro Bologna and Bernardo Tellini}

\date{ Instituto de Alta Investigaci\'{o}n,
Universidad de
  Tarapac\'{a}-Casilla 7-D Arica, Chile\\
Department of Energy and Systems Engineering, University of Pisa,
Largo L. Lazzarino, I-56122 Pisa, Italy.}\maketitle



\abstract {In this paper we analytically estimate the magnetic
field scale of planets with physical core conditions similar to
that of Earth from a statistical point of view. We evaluate the
magnetic field on the basis of the physical parameters of the
center of the planet, such as density, temperature, and core size.
We look at the contribution of the Peltier-Seebeck effect on the
magnetic field, showing that an electrical thermal current can
exist in a rotating fluid sphere. Finally, we apply our
calculations to Earth and Jupiter. In each case we show that the
thermal generation of currents leads to a magnetic field scale
comparable to the observed fields of the two planets.}

\section{Introduction}
The Earth's magnetic field is a fascinating problem that has been
faced by many authors and the reader can find rich literature on
the topic. For many years the intuitive idea that the magnetic
field is generated by heavy fluid in the center of Earth subjected
to the rotational motion of our planet, has been conjectured. Many
numerical works have started to shed light on the possible
mechanism of the generation of Earth's magnetic field. The basic
model for the generation of Earth's magnetic field or of other
planets, is based upon the dynamo effect of a turbulent convection
in rotating fluids. This idea has received much attention in the
past few years and many numerical studies based on the dynamo
model attempted to reproduce some of the main properties of the
magnetism of celestial bodies~\cite{gla,kuang,busse,kono2}, among
them the phenomenon of magnetic field reversal. Magnetic field
reversal, the phenomenon for which the positions of magnetic north
and magnetic south are interchanged, is another important feature
of the terrestrial magnetic field that has been studied
intensively, and recently a similar phenomenon has been reproduced
in the laboratory~\cite{ber}.

In a previous work~\cite{bt} the authors showed that a magnetic
field can be generated in the laminar region of a fluid velocity
under the condition~$\rho=\eta\sigma\mu$, but such a condition is
far from the usual condition of the celestial body and, more
important, its magnitude would be of the order of~$B\sim\Omega R
\sqrt{\rho \mu_0}$ where~$\Omega$ is the rotational velocity,~$R$
is the radius of the outer core, and~$\rho$ is the density of the
fluid in the outer core of Earth~\cite{core}. Inserting
Earth's parameters would imply an intensity field of~$B\sim 10^5~$
gauss that is very far from Earth's actual magnetic field value.
Independent from the path that the above value has been
obtained, it is a fact that the magnetic field scale~$B\sim\Omega
R \sqrt{\rho \mu_0}$ can be deduced by magnetohydrodynamic
equations. This is an indication that the Earth's magnetic field
could have a different origin that is not strictly dynamic, since the
value appears to be too high when compared with the observed
field. The above discussion leads us to the following question:
Does there exist for Earth a magnetic field scale as a function
of the physical system parameters? We shall show that
there exists such a characteristic magnetic field with an
intensity that is very close to the actual value of Earth's
field.

\section{Magnetohydrodynamic equations}\label{sec2}

Let us consider the set of equations for a plasma with finite
conductivity and constant density~\cite{landau8,jacky}

\begin{eqnarray}\label{idro1}
 \rho\left[\frac{\partial}{\partial t}\mathbf{v}+
(\mathbf{v}\cdot \texttt{{\boldmath$\nabla$}}) \mathbf{v
}\right]&=&-\texttt{{\boldmath$\nabla$}} P+
\left[\texttt{{\boldmath$\nabla$}}\times \mathbf{H}\right]\times
\mathbf{B}+\mathbf{f}+
\texttt{{\boldmath$\sigma$}}\\
\label{idro2}
\texttt{{\boldmath$\nabla$}}  \cdot\mathbf{v }&=&0\\
\label{idro3} \frac{\partial \mathbf{B}}{\partial t} &=&
\texttt{{\boldmath$\nabla$}} \times \left[\mathbf{v }\times
\mathbf{B}\right]
+\frac{1}{\mu\sigma}\nabla^{2}\mathbf{B}\\
\label{idro4} \texttt{{\boldmath$\nabla$}} \cdot\mathbf{B }&=&0,
\end{eqnarray}
where~$\mathbf{v}$ is the flow velocity,~$\mathbf{H}$ is the
magnetic field related to the magnetic induction~$\mathbf{B}$ via the
relation~$\mathbf{B}=\mu \mathbf{H}$,~$P$ is the pressure of the
gas, and~$\rho$ is the mass density. The gravity force
density,~$\mathbf{f}$, takes the form~$\mathbf{f}=\rho
\texttt{{\boldmath$\nabla$}}\psi$ . The vector~$
\texttt{{\boldmath$\sigma$}}$ is defined through its components as
follows~\cite{landau6}

\begin{equation}\label{land1}
 \sigma_i=\frac{\partial \sigma'_{ik}}{\partial x_k},
\,\,\,\,\, \sigma'_{ik}=\eta\left(\frac{\partial v_{i}}{\partial
x_k}+\frac{\partial v_{k}}{\partial x_i} \right)
\end{equation}
where~$\sigma'_{ik}$ is the viscous stress tensor, and~$\eta$ is
the coefficient of viscosity which is assumed constant. We also
used the convention of dropping the symbol of sum for the repeated
indexes. The current density~$\mathbf{J}$ is given by the
constitutive relation~$\mathbf{J}=\sigma
(\mathbf{E}+\mathbf{v\times B})$ where~$\sigma$ is the electrical
conductivity of the fluid.

The dynamic system has to be implemented using the equation of heat
transfer in magnetohydrodynamics~\cite{landau8}

\begin{equation}\label{land_heat}
\rho c_p\left(\frac{\partial}{\partial t}T+ \mathbf{v } \cdot
\texttt{{\boldmath$\nabla$}} T\right)= \sigma'_{ik} \frac{\partial
v_{i}}{\partial x_k}+\kappa\nabla^2 T+\frac{J^2}{\sigma}+Q
\end{equation}
where~$c_p$ is the specific heat at constant pressure,~$\kappa$ is
the thermal conductivity,~$T$ is the temperature of the fluid,
and~$Q$ is the quantity of heat generated by external sources of
heat contained in a unit volume of the fluid per unit time.

Let us define a dimensionless velocity~$\mathbf{u}=
\mathbf{v}/(\bar{\Omega} R)$, a dimensionless operator
$\texttt{{\boldmath$\nabla$}}'=R\texttt{{\boldmath$\nabla$}}$, and
a dimensionless time~$\tau=\bar{\Omega} t$. Combining
(\ref{idro1}) and (\ref{land1}) we can rewrite the hydromagnetics
equations as

\begin{eqnarray}\label{idro1_b}
 \frac{\partial}{\partial \tau}\mathbf{u}+
(\mathbf{u}\cdot \texttt{{\boldmath$\nabla$}})
\mathbf{u}&=&-\frac{1}{R_P}\texttt{{\boldmath$\nabla$}}
(p+\bar{\psi})+
\frac{1}{R_B}\left[\texttt{{\boldmath$\nabla$}}\times
\mathbf{b}\right]\times
\mathbf{b}+\frac{1}{Re}\texttt{{\boldmath$\nabla$}}^{2}\mathbf{u}
\\
\label{idro3_b} \texttt{{\boldmath$\nabla$}} \cdot \mathbf{u}&=&0
\\
\label{idro2_b} \frac{\partial \mathbf{b}}{\partial \tau} &=&
\texttt{{\boldmath$\nabla$}} \times \left[\mathbf{u }\times \mathbf{b}\right]+
\frac{1}{R_M}\nabla^{2}\mathbf{b}\\
\label{idro4_b} \texttt{{\boldmath$\nabla$}} \cdot\mathbf{b}&=&0
\end{eqnarray}
where we defined the dimensionless quantities~$p=
 P/P_0$,~$\bar{\psi}= \rho\psi/P_0$,~$
 \mathbf{b}= \mathbf{B}/B_0$
with~$P_0$ as a characteristic pressure, and~$B_0$ as a
characteristic induction field. For brevity we redefined
$\texttt{{\boldmath$\nabla$}}'$ as~$\texttt{{\boldmath$\nabla$}}$.
Finally we defined the numbers

$$
  Re= \frac{\rho\bar{\Omega} R^2}{\eta},\,\,\,\,\,\,\,
 R_P=\frac{\rho\bar{\Omega}^2 R^2}{P_0}
 ,\,\,\,\,\,\,\,R_B=\mu\frac{\rho\bar{\Omega}^2 R^2}{B_0^2},
 \,\,\,\,\,\,\,R_M= \sigma \mu
\bar{\Omega} R^2
$$
where~$Re$ is the Reynolds number,~$R_B$ the magnetic force number,
and~$ R_M= \sigma \mu \bar{\Omega} R^2$ is the Reynolds magnetic
number.

As many authors have pointed out the analytical solution of this
equation is a very hard task and only a few exact or approximate
cases are known~(see for example~\cite{libro} for a review). Our
main aim is to find an estimation for~$B_0$ as a function of the
physical parameters of the center of Earth (such as density,
temperature, etc.) without necessarily solving the system
(\ref{idro1_b})-(\ref{idro4_b}).

\section{Remarks on field velocity components for a rotating sphere}
In this section we will show that for a rotating sphere, in general
the~$\phi$ component of the velocity is not enough to
describe a time dependent motion, with it being understood
that~$v_{\phi}=\Omega r \sin\theta$, where~$\Omega$ is the angular
velocity, is an exact solution. Note that in Ref.~\cite{kerr}
is shown evidence that the inner core of the Earth may spin
faster than the rest of the planet so that the above exact
solution does not hold for the fluid motion of the terrestrial
core. This conclusion implies that the temperature distribution
can not be a radial distribution due to the fact that
Eq.~(\ref{land_heat}) is, in general, coupled with the velocity field.
To enforce this statement we shall show that in general a rotating
fluid sphere, as previously stated, can not be described by the~$\phi$
component of velocity. Let us then assume that a sphere starts
to rotate with $v_{\phi}(t)$ such that~$v_{\phi}(0)=0$
and~$v_{\theta}=v_{r}=0$. We also assume that for symmetry all
physical variables do not depend on~$\phi$. Writing only the
hydrodynamic part of the set of
equations~(\ref{idro1})~-~(\ref{idro4}), i.e. setting
$\mathbf{B}=0$, we obtain

\begin{eqnarray}\label{vf1}
 \cot\theta \frac{v_{\phi}^2}{r}&=&
 \frac{1}{\rho r}\frac{\partial P}{\partial\theta}
\\
\label{vf2} \frac{v_{\phi}^2}{r}&=&\frac{1}{\rho }\frac{\partial P
}{\partial r}
\\
\label{vf3} \frac{\partial v_{\phi}}{\partial t} &=&
  \frac{\eta}{\rho}\nabla^{2}v_{\phi}.
\end{eqnarray}
Combining Eqs.~(\ref{vf1}) and~(\ref{vf2}) we obtain

\begin{equation}\label{velphi}
\cot\theta \frac{\partial }{\partial r}v_{\phi}^2  =\frac{1}{r
}\frac{\partial }{\partial \theta}v_{\phi}^2.
\end{equation}
The above equation is satisfied by a function~$v_{\phi}(r,\theta)$
of the form~$v_{\phi}(r,\theta)=v_{\phi}(r\sin\theta)$. Using this
result we may rewrite Eq.~(\ref{vf3}) as

\begin{equation}\label{velphi2}
\frac{\partial}{\partial t} v_{\phi}(t,x) =
\frac{\eta}{\rho}\left(\frac{\partial^{2}}{\partial
x^{2}}+\frac{1}{x} \frac{\partial  }{\partial
x}-\frac{1}{x^2}\right)v_{\phi}(t,x),\,\,\,\,x\equiv r\sin\theta.
\end{equation}
The general solution of the above equation, via Laplace transform
and with the condition~$ v_{\phi}(0,x)=0$, is

\begin{equation}\label{velphi3}
 \hat{ v}_{\phi}(s,x) =a
I_1\left(\frac{x \sqrt{s}}{\sqrt{D}}\right)+b K_1\left(\frac{x
\sqrt{s}}{\sqrt{D}}\right)
\end{equation}
where, by definition,~$ \hat{
v}_{\phi}(s,x)=\int_{0}^{\infty}\exp[-s t]v_{\phi}(t,x)dt$ and
$D=\eta/\rho$. Assuming a finite solution for~$r\to 0$ and
consequently~$x\to 0$, then~$b=0$.  It is evident by the form of
Eq.~(\ref{velphi3}) that the boundary condition for the velocity,
i.e.~$\hat{v}_{\phi}(s,R\sin\theta)=a I_1\left( R\sin\theta
\sqrt{s/D}\right)$ where~$R$ is the sphere radius, can be
satisfied only by a restricted class of functions. One could
consider adding another component, for example the radial
component~$v_r$, to describe matter falling in the center. But the
independency of the dynamic by the angular variable~$\phi$ implies
that the~$\theta$ component of the velocity~$v_{\theta}$ also
has to be considered. This fact comes directly from the continuity
equation. Let us assume that there is another component of  velocity,
the radial component~$v_r$. The
continuity equation is

\begin{equation}\label{idro2_bis}
\frac{1}{r^2}\frac{\partial}{\partial r}(r^2 v_r)=0.
\end{equation}
The solution~$v_r=f(\theta)/r^2$ diverges at the origin and
can not vanish on the surface of the sphere~$r=R$. To avoid this
inconsistency we are forced to add~$v_{\theta}$ to the flow. From
this we can infer that from the early stages of Earth's formation to
the present, the velocity of the fluid could not be
described by only one component of the velocity vector. This
analytical conclusion is in agreement with the numerical works
presented in the references.

\section{Thermal generation of magnetic field}
Looking at Eq.~(\ref{idro3}), or Eq.~(\ref{idro3_b}), we notice
that is a diffusive-like equation. For the time
evolution of~$\mathbf{B}$ it is important to give an estimation of
its initial value. It is accepted that the Earth's core is made mainly of
iron with a solid inner core the size~$10^3$ km and an outer core of
liquid about~$2\times 10^3$ km thick~\cite{core}.
The temperature distribution of the core is not uniform and it
ranges from approximatively~$10^4$~$^{\circ}$K at the very center
to~$10^3$~$^{\circ}$K at the surface of the outer core. The non
uniform temperature can generate a contribution to the electrical
current that is proportional to the gradient of the temperature,
known as the Peltier$-$Seebeck effect, so that we can write
the total current as~\cite{landau10}

\begin{equation}\label{curr_therm}
\mathbf{J}=\sigma\left[\mathbf{E}+\mathbf{v}\times \mathbf{B}
-\alpha(T)\texttt{{\boldmath$\nabla$}}T\right].
\end{equation}
Note that Eq.~(\ref{idro3}) does not
change if~$\alpha(T)\texttt{{\boldmath$\nabla$}}T$ can be
written as a gradient of a function. To evaluate the coefficient~$\alpha$,
we have to consider
the fact that the density of either the solid inner core, or the fluid
outer core, is such that the electrons can be
considered a degenerate Fermi's gas. Indeed, according to the
authors of Ref.~\cite{nat} the core density is of the order of
$10^4\,\, \mathrm{Kg\, m}^{-3}$. This implies that the Fermi
energy of the electrons of the Earth's core is
\begin{equation}\label{fermi}
\varepsilon_F=\frac{\hbar^{2}}{2m}\left(3\pi^2\frac{N}{V}\right)^{2/3}\approx
 2\times 10^{-18}\,\,\mathrm{Joule}
\end{equation}
corresponding to a Fermi temperature~$T_F\approx 1.4\times 10^5\,^{\circ}$K.
This is at least one order of magnitude higher than the Earth's
core temperature therefore
justifying the degenerate Fermi's gas approximation.
Quantum calculations show that~\cite{landau10}

\begin{equation}\label{alpha}
\mid\alpha(T)\mid\sim k_B \frac{k_B T}{e \varepsilon_F}
\end{equation}
where~$k_B$ is the Boltzmann constant and~$e$ is the electron
charge. Let us consider a very simplified model of the early stage's of Earth's
formation. Models suggest that Earth's core was completely molten~\cite{buf}.
Since the temperature change
of the core has a time scale on the order of
Earth's age~\cite{nat3} we infer that whatever is the contribution
from the thermal term in Eq.~(\ref{curr_therm}) this
contribution still holds today with the same order of magnitude. A widely
accepted estimation of the core temperature is
approximatively~$8\times 10^3$~$^{\circ}$K for the inner core
and~$4\times 10^3$~$^{\circ}$K for the outer core~(see for
example~\cite{nat,poir,alfe}). Let us consider the
contribution to the magnetic field due to thermal term.
From Maxwell's equation we obtain

$$\texttt{{\boldmath$\nabla$}}\times \mathbf{B}=\mu\sigma
\mathbf{J}= -\mu_0\sigma
\alpha(T)\texttt{{\boldmath$\nabla$}}T.$$ In general, the
temperature distribution in time and space is coupled with the
velocity field, Eq.~(\ref{land_heat}), and as shown in the
previous section, all components of the velocity are present so
that the temperature distribution can not be only radial. Note
also that we use~$\mu_0$ as the value of the magnetic permeability
since at such temperature we assume that there is no
magnetization. We can deduce the field scale via the relation

\begin{equation}\label{fields}
\frac{B}{R}\approx \mu_0\sigma \alpha(T)\frac{\Delta T}{R}
\end{equation}
where~$\Delta T$ is the difference in temperature, and~$R$ is the
length scale of the system. We obtain the scale strength of the
thermal magnetic field

\begin{equation}\label{best}
B_{T}=\mu_0\sigma \frac{k_B T_c}{e \varepsilon_F} k_B\Delta T.
\end{equation}
Using the value of temperature~$T_c\approx 8\times 10^3$
$^{\circ}$K,~$\Delta T\approx 4\times 10^3$~$^{\circ}$K,
and~$\sigma\approx 10^5\,\,\mathrm{S\, m}^{-1}$~\cite{sta} we
obtain the numerical value for the strength of the core of Earth 's
magnetic field

\begin{equation}\label{best2}
B_{T}\approx 0.0024\,\,\textrm{tesla}=24 \,\,\textrm{gauss}.
\end{equation}
The estimated strength of the core of Earth 's
magnetic field is
approximatively~$B_{est}\sim 25$ gauss~\cite{nat2} which is very
close to the analytical value given by Eq.~(\ref{best}). We note
that selecting different values for the temperature, according to
the different models present in the literature, the value of the
field would of course change consequently, but the scale magnitude
remains of the order of tens of gauss. The thermal current
gives a strong (if not total) contribution  to the magnetic field. It
is worthy to stress that in principle the only phenomenological parameter,
i.e. the conductivity~$\sigma$, could be evaluated using
quantum mechanics~\cite{nat4} so that we can conclude that the
field given by the expression (\ref{best}) may be written in terms
of fundamental constant and physical parameters of the
system such as density~$N/V$ and temperature~$T$. Note that
dependence on the radius~$R$ of the fluid region is implicit in
the dependence of the temperature on~$R$. This is the main reason
why we kept explicit the temperature difference~$\Delta T$ in
Eq.~(\ref{best}). Once we clarify this, we can rewrite more
concisely Eq.~(\ref{best}) as

\begin{equation}\label{best_bis}
B_{T}=\mu_0\sigma \frac{\left(k_B \bar{T}\right)^2}{e
\varepsilon_F}
\end{equation}
where~$\bar{T}$ can be taken, for example, as the average temperature of the
fluid region. We conclude that Eq.~(\ref{best_bis}) represents the
scale of the strength of the magnetic field of celestial bodies
with an Earth-like physical condition for the core, from a statistical point of view.

\section{Jupiter's magnetic field estimation}
In principle we can apply the ideas of the previous section to
other celestial bodies, particularly in our solar system. The
main difficulty with this is the scarcity of information about
the physical internal condition of other planets, although we can
make some general considerations. For example, Mercury and Mars are
quite smaller than Earth. This fact surely contributes to a
faster cooling of their interiors so we can expect that the cores
of these planets are no longer in the fluid state. In fact the two planets
have a very weak magnetic field.

Venus does not have a magnetic field~\cite{nel} and there are
several possible explanations for this. Venus is a planet very
similar to Earth in dimension but it does not exhibit volcanic
activity, and this could imply a cold core. Venus has a very slow
rotational motion compared to Earth, and a rotational
motion is considered to play a crucial role for the terrestrial
magnetic field. Also we should consider the possibility that Venus
could be in a reversal phase.

Jupiter is a good candidate to test our model. Even though little is known about the planet,
its internal structure has been modeled by several authors~(see for
example~\cite{hub,nad,yagi}) and the physical information is enough to allow a rough
estimation of the scale of its magnetic field using
Eq.~(\ref{best}). The electrical conductivity is~$\sigma\approx
10^5\,\, \mathrm{S\,m^{-1}}$~\cite{yagi}, its temperature ranges
from $T\approx 2\times 10^4$$^{\circ}$K for the core boundary to
$T\approx 10^4$$^{\circ}$K for the metallic hydrogen boundary, and
the estimated density of the metallic hydrogen is~$\rho\approx
4\times 10^3\,\,\mathrm{Kg\,m}^{-3}$~\cite{hub,nad}. Consequently
the Fermi's energy takes the value~$\varepsilon_F\approx 10^{-17}$
Joule corresponding  to a Fermi temperature~$T_F\approx 7\times
10^5~$~$^{\circ}$K so that we can apply the Fermi statistic for
the electrons in the metallic region. Plugging these values into
Eq.~(\ref{best}) we obtain for the magnetic field of Jupiter an
estimation of its strength in the metallic hydrogen
region~$B_{J}\approx 30$ gauss. Taking into account that this region
extends for a fraction, ranging from~$0.7$ to $0.78$ times the Jupiter
radius, we obtain for the surface value of the magnetic field a range from~$B_{JS}\approx
B_{J}(0.7)^3\approx 10$ gauss to~$B_{JS}\approx
B_{J}(0.78)^3\approx 14$ gauss. This is in agreement with the scale of the observed
values~\cite{smith}. As said for Earth, according to several models,
we can change the values of the parameters, but the
strength of the field would still be of the order of the observed
field.
\section{Conclusion}
We provided an analytical estimation of the magnetic field scale
of planets with physical core conditions similar to that of
Earth from a statistical point of view.
The magnetic field strength was
evaluated directly from the
physical parameters of the center of the planet, considering density,
temperature and core's size. We showed that an electrical current
generated by a thermal gradient can exist in a rotating fluid
sphere and can give an important contribution to the magnetic field.
Our conjecture was supported by estimating the magnetic field
strengths of Earth and Jupiter that were in agreement with the
observed magnetic field intensity of the two planets.

\section{Acknowledgments}
M.B. acknowledges financial support from FONDECYT project no
1110231. The authors thank Catherine Beeker for her editorial
contribution.

\end{document}